\documentclass[lettersize,journal]{IEEEtran}
\usepackage{amsmath,amsfonts}
\usepackage{algorithmic}
\usepackage{array}
\usepackage[caption=false,font=normalsize,labelfont=sf,textfont=sf]{subfig}
\usepackage{textcomp}
\usepackage{stfloats}
\usepackage{url}
\usepackage{verbatim}
\usepackage{graphicx}
\usepackage{xcolor}
\def\BibTeX{{\rm B\kern-.05em{\sc i\kern-.025em b}\kern-.08em
    T\kern-.1667em\lower.7ex\hbox{E}\kern-.125emX}}
\usepackage{balance}
\begin{document}
\title{Enhanced Direction-Sensing Methods and Performance Analysis in Low-Altitude Wireless Network via a Rotating Antenna Array}
\author{Jinbing Jiang, Feng Shu, Minghao Chen, Jiatong Bai, Maolin Li, Yan Wang and Jiangzhou Wang 
\thanks{Manuscript created October, 2020; This work was developed by the IEEE Publication Technology Department. This work is distributed under the \LaTeX \ Project Public License (LPPL) ( http://www.latex-project.org/ ) version 1.3. A copy of the LPPL, version 1.3, is included in the base \LaTeX \ documentation of all distributions of \LaTeX \ released 2003/12/01 or later. The opinions expressed here are entirely that of the author. No warranty is expressed or implied. User assumes all risk.}}

\markboth{Journal of \LaTeX\ Class Files,~Vol.~18, No.~9, September~2020}%
{How to Use the IEEEtran \LaTeX \ Templates}

\maketitle

\begin{abstract}
Due to the directive property of each antenna element, the received signal power can be severely attenuated when the emitter deviates from the array boresight, which will lead to a severe degradation in  sensing performance along the corresponding direction. 
Although existing rotatable array sensing methods such as recursive rotation root multiple signal classification (RR-Root-MUSIC) can mitigate this issue by iteratively rotating and sensing, several mechanical rotations and repeated eigendecomposition operations are required to yield a high computational complexity and low time-efficiency. 
To address this problem, a pre-rotation initialization with receive power as a rule is proposed to significantly reduce the computational complexity and improve the time-efficiency. Using this idea, a low-complexity enhanced direction-sensing framework with pre-rotation initialization and iterative greedy spatial-spectrum search (PRI-IGSS) is developed with three stages: (1) the normal vector of array is rotated to a set of candidates to find the optimal direction with the maximum sensing energy with the corresponding direction-of-arrival (DOA) value computed by the Root-MUSIC algorithm; (2) the array is mechanically rotated to the initial estimated direction and kept fixed; (3) an iterative greedy spatial-spectrum search or receiving beamforming method, motivated by reinforcement learning, is designed with a reduced search range and  making a summation of all previous sampling variance matrices and the current one is adopted to provide an increasing performance gain as the iteration process continues. To assess the performance of the proposed method, the corresponding Cramer-Rao lower bound (CRLB) is derived with a simplified rotation model. Simulation results demonstrate that the proposed PRI-IGSS method performs much better than RR-Root-MUSIC and achieves the CRLB in term of mean squared error due to the fact there is no sample accumulation for the latter. Compared to the latter, the former also makes a significant reduction in computational complexity.
\end{abstract}


\section{Introduction}
\label{sec1}
With the rapid development of the low-altitude economy, unmanned aerial vehicles (UAVs) have gradually become an indispensable part of future wireless networks, enabling a wide range of applications from logistics distribution and emergency rescue to aerial inspection \cite{Low-Altitude-apply1, Low-Altitude-apply2, Low-Altitude, Low-Altitude-Sensing-demand}. \textcolor{black}{To support these low-altitude applications, accurate UAV spatial localization and beam alignment are essential, both of which rely on precise angular information. Therefore, direction-of-arrival (DOA) estimation plays a fundamental role in low-altitude UAV sensing and communication, especially in millimeter-wave systems where narrow-beam transmission makes beam alignment highly sensitive to DOA accuracy} \cite{UAV-ISAC-YY, UAV-XZY}. It is well known that various DOA estimation techniques have been extensively investigated, primarily including conventional beamforming, maximum likelihood estimation, and high-resolution subspace-based algorithms. \textcolor{black}{However, their application to low-altitude UAV network is still challenging due to the computational complexity  and the physical received-power limitations.}

\textcolor{black}{To achieve cost-effective and low-complexity DOA estimation for UAV sensing, hybrid analog-digital (HAD) and heterogeneous HAD have been widely studied \cite{LC-DOA-SF, HAD-SF}. Meanwhile, several low-complexity DOA estimators, such as rapid phase ambiguity elimination methods \cite{RPAE-DOA-ZXC}, multi-modal deep-learning-aided estimators \cite{HAD-BJT}, and root multiple signal classification (Root-MUSIC) based algorithms \cite{LC-DOA-CYW}, have been proposed to reduce the computational complexity caused by covariance estimation, eigenvalue decomposition, or spectral search. However, these existing studies focus on receiver architecture design or signal processing complexity reduction under a given received signal quality.  In low-altitude UAV sensing, when the UAV deviates from the array boresight or falls within the null region of a fixed directional antenna array, the effective received signal-to-noise ratio (SNR) can drop sharply, leading to severe DOA estimation performance degradation due to directional antenna pattern \cite{DA-UAV-DWK,2003Antenna}. Therefore, purely algorithmic complexity reduction cannot address the antenna gain loss caused by directional antenna pattern.}

\textcolor{black}{This physical limitation has motivated the investigation of reconfigurable antenna technologies and reconfigurable propagation environments for improving the received signal quality. Among these technologies, movable antenna (MA) has been proposed as a position-domain reconfigurable technology, where antenna positions are adjusted over a continuous spatial region to provide additional spatial degrees of freedom for communication and sensing \cite{R3,6D-MA}. For MA communications, a compressed-sensing-based channel estimation method was developed in \cite{R1}, where the angular-domain field-response information is estimated from finite channel measurements to reconstruct the channel response over the antenna movement region. It was further shown in \cite{R2} that antenna position optimization can significantly enhance wireless sensing performance. Recent studies have also extended MA to various communication and sensing scenarios, such as interference suppression via null steering \cite{MA-EB-ZLP}, UAV data collection \cite{Low-Altitude-MA}, near-field communications \cite{R4}, hybrid beamforming \cite{R5}, discrete antenna positioning under imperfect CSI \cite{R6}, and energy-efficiency optimization with movement-related time and energy costs \cite{R7}.
	Closely related to MA, fluid antenna systems (FASs) have been introduced as a flexible antenna paradigm, where the antenna position, shape, or physical structure can be dynamically reconfigured to adapt to wireless channel conditions. By exploiting this reconfigurability, additional spatial degrees of freedom can be obtained to enhance channel diversity, connectivity, and interference suppression \cite{FA-NWK, FA-KKW2026}. In parallel, intelligent reflecting surfaces (IRSs) have been proposed as propagation-domain reconfigurable technologies. By adjusting passive or active reflecting elements, the wireless propagation environment can be reshaped. IRSs have been widely applied to improve communication performance \cite{RIS-HCW}, physical-layer security \cite{DM-IRS-SF}, and sensing capability \cite{IRS-Sensing-WQQ} through channel reconfiguration and joint beamforming design.}

\textcolor{black}{However, the directional antenna gain loss considered in this paper is essentially an element-orientation issue, which cannot be directly resolved by position-domain or propagation-domain reconfigurability.   
	For MA and FAS, antenna repositioning can improve the spatial channel condition, but it cannot change the gain pattern or boresight direction of each directional receiving antenna. 
	Therefore, when the UAV direction is far away from the antenna boresight, the antenna gain may still be severely attenuated if the element boresight remains unchanged. 
	Similarly, IRS can reshape the propagation channel between the transmitter and receiver, but it does not alter the intrinsic gain characteristics of the receiving elements \cite{IRS-DM-DRE,DoF-JJB}. 
	As a result, when the UAV falls within a low-gain region, such as a null region or the backside of a fixed directional receiving array, the effective received SNR can be severely degraded.  This directivity-induced SNR loss cannot be directly eliminated without reorienting the directional receiving elements.}

\textcolor{black}{To address this orientation mismatch between the antenna boresight and the UAV direction, the orientation-domain reconfigurability is introduced by a rotatable antenna array \cite{RA-2ZBX}. 
	By rotating the array, the normal direction of the array can be aligned with the UAV direction. As a result, the UAV signal can be received from the high-gain direction of the directional elements, and the effective received SNR can be improved without complex internal array reconfiguration \cite{RA-ZBX}.
	For instance, rotatable antenna arrays have been shown to improve hybrid precoding performance for UAV millimeter-wave communications by mechanically aligning the antenna boresight with the desired direction \cite{RAA-UAV}. 
	Therefore, rotatable antenna array is complementary to MA/FAS and IRS. While MA/FAS and IRS mainly exploit position-domain and propagation-domain reconfigurability, respectively, the rotatable antenna array directly manipulates the orientation of directional elements and is particularly suitable for mitigating off-boresight gain loss in UAV direction sensing.}

\textcolor{black}{While rotatable antenna provides a direct channel gain advantage, existing rotatable antenna array-aided DOA estimation methods still suffer from computational overhead. 
	For example, recursive rotation Root-MUSIC (RR-Root-MUSIC) \cite{Jiang2025RotatableAD} relies on repeated repeated eigenvalue decompositions (EVDs) to refine the DOA estimate, which may lead to high sensing latency and computational complexity. 
	To address this problem, a low-complexity pre-rotation initialization and iterative greedy spatial-spectrum search (PRI-IGSS) method is proposed in this paper. In the proposed method, rotating array is used to increase the directional antenna gain and thereby improve the received SNR, while the final DOA is obtained from Root-MUSIC initialization and IGSS-based refinement. It is worth noting that the proposed method is different from conventional mechanically scanned radar (MSR). In MSR, the antenna beam is mechanically swept over an angular sector, and the mechanical pointing angle with the strongest received response is usually used to infer the target direction. In contrast, the proposed method is a subspace-based DOA estimation assisted by a rotatable antenna array. The array rotation is used to compensate for the off-boresight gain loss of directional receiving elements and to improve the effective received SNR. The final DOA is obtained by Root-MUSIC initialization and IGSS-based refinement. Then, both the sensing signal quality and the search efficiency can be improved, which makes the proposed method suitable for low-complexity UAV direction sensing.} The main contributions of this paper can be summarized as follows:
\begin{enumerate} 
	\item To tackle the severe signal attenuation caused by the directive pattern of antenna elements, a rotatable array direction-sensing system model is established. By rotating the array toward the target, it can effectively ease the problem of sensing performance degradation particularly when the emitter deviates from the array boresight. Furthermore, based on this model, the corresponding Cramer-Rao lower bound (CRLB) is derived by virtue of statistical theory and matrix analysis, which provides a theoretical benchmark to evaluate the following direction-sensing methods.
	
	\item To achieve a high-precision direction-sensing performance with low-complexity, a PRI is proposed. First, the array is rotated to a set of candidate angles, which is equi-spaced rotation or sampling along elevation and azimuth to find the optimal direction with the maximum received power, and an initial direction is estimated by the Root-MUSIC algorithm. Subsequently, the array is rotated to this sensed direction and kept to be fixed. Finally,  an IGSS is iteratively performed as the search range is reduced gradually, and the sensing precision grows gradually  with a continuous sampled signal accumulation process, which sum up to the current sampling variance matrix including all previous ones in the current round. Here, CRLB is utilized to design the search step-size. The proposed method makes a great sensing performance improvement over the conventional RR-Root-MUSIC and proposed PRI during the summation of sampling variance matrices. Their computational complexities are in decreasing order as follows: RR-Root-MUSIC, PRI-IGSS, and PRI.

	\item In accordance with simulation results, the proposed PRI-IGSS method can achieve the corresponding CRLB. More important, by exploiting a continuous sample accumulation mechanism, the performance of the proposed method is significantly improved compared to the traditional non-memory RR-Root-MUSIC. Specifically, across the entire SNR and angular domains, an achievable performance gain of approximately one order of magnitude over RR-Root-MUSIC is consistently maintained. Furthermore, in extreme off-boresight scenarios, e.g., $\theta > 80^o$, the proposed PRI and PRI-IGSS exhibit tremendous spatial robustness; while conventional FA (fixed array)-Root-MUSIC suffers severe degradation, resulting in a massive error gap of up to three and four orders of magnitude, the proposed PRI-IGSS remains highly reliable and accurate. Additionally, an optimal pre-rotation configuration of $Q$ = 3 is verified to perfectly balance the array high-gain coverage and the system overhead.
	
\end{enumerate} 

The remainder of this research is organized as follows. The system model of rotatable array is described in Section II. In Section III, the corresponding CRLB is also derived. In Section IV, two methods, PRI and PRI-\textcolor{black}{IGSS}, are proposed. Moreover, section V presents the experimental results, with conclusions provided in Section VII.

\section{System Model}

\begin{figure}
	\centering
	\includegraphics[width=0.7\linewidth]{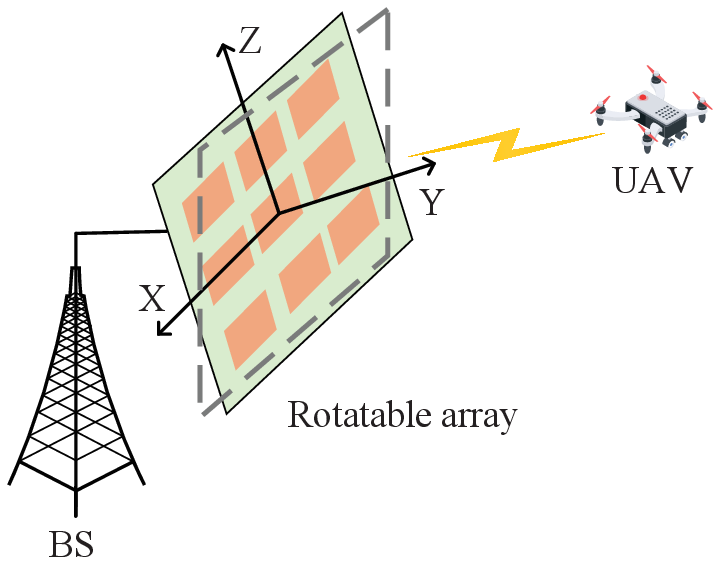}
	\caption{rotatable array system for low-altitude communication network.}
	\label{fig:ra}
\end{figure}
As shown in Fig. \ref{fig:ra}, a rotatable array system for low-altitude communication network is considered where UAV is equipped with a single isotropic fixed antenna. The base station (BS) is equipped with a rotatable uniform planar array (UPA), which is composed of directional antennas. Without loss of generality, the array is deployed on the $x$-$z$ plane of a three-dimensional (3D) Cartesian coordinate system with the origin as the center, and its size is $M\times N$, where $M$ and $N$ represent the number of antennas along the $x$-axis and $z$-axis, respectively.
The separations between adjacent antennas in two axes are denoted by $d_x$ and $d_z$,  respectively. Therefore, the whole UPA size can be expressed as $Md_x\times Nd_z$.

For convenience, assuming that both $M$ and $N$ are odd numbers, the $(m,n)$-th antenna is located at the $m$-th column ($m=0,...,M-1$) and $n$-th row ($n=0,...,N-1$) on the UPA, which can be written as
\begin{align}
	\mathbf{p}_{m,n}=[x_m,0,z_n  ]^T,
\end{align}	
where $x_m= (m-\frac{M-1}{2})d_x$ and $z_n=(n-\frac{N-1}{2})d_z$.

\begin{figure}[!t]
	\centering
	\subfloat[Initial orientation]{
		\includegraphics[width=0.2\textwidth]{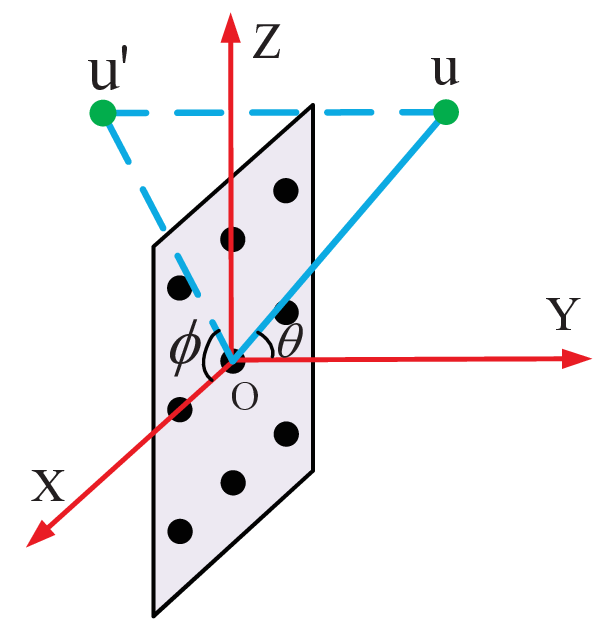}%
		\label{subfig:xyz-a}
	}
	\hfil
	\subfloat[Rotated orientation]{
		\includegraphics[width=0.2\textwidth]{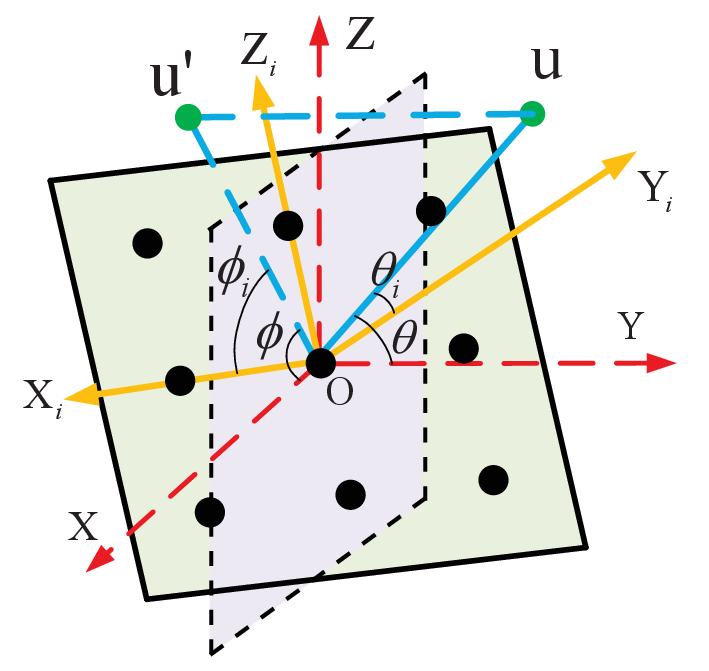}%
		\label{subfig:xyz-b}
	}
	\caption{Illustration of the geometric relationship for the rotatable array and emitter.}
	\label{fig:xyz}
\end{figure}

Fig.~\ref{fig:xyz} illustrates the geometric relationships for the rotatable array and emitter. The angle $\theta$ between the emitter direction and the $y$-axis, and $\phi$ between the projection of the emitter direction onto the $x$-$z$ plane and the $x$-axis, are denoted as the elevation and azimuth angles of the UAV with respect to the origin of the coordinate system, respectively. Accordingly, the unit direction vector of the emitter is shown as
\begin{align}\label{OS}
	\mathbf{u}&=[u_x,u_y,u_z]^T\nonumber\\
	&=[\sin\theta\cos\phi,\cos\theta,\sin\theta\sin\phi]^T,
\end{align}
where $\theta\in [0,\pi]$ and $\phi\in [-\pi,\pi]$.

Hence, the array manifold is given by
\begin{align}
	\mathbf{a}(\theta,\phi)=\mathbf{a}_z(u_z) \otimes \mathbf{a}_x(u_x),
\end{align}
where
\begin{equation}
	\mathbf{a}_z(u_z) = 
	\begin{bmatrix}
		e^{j\frac{2\pi}{\lambda} z_0  u_z} \\
		\cdots  \\
		e^{j\frac{2\pi}{\lambda}z_n  u_z}  \\
		\cdots  \\
		e^{j\frac{2\pi}{\lambda} z_{N-1}   u_z}
	\end{bmatrix},
	\mathbf{a}_x(u_x) = 
	\begin{bmatrix}
		e^{j\frac{2\pi}{\lambda} x_0  u_x} \\
		\cdots  \\
		e^{j\frac{2\pi}{\lambda}x_m u_x}  \\
		\cdots  \\
		e^{j\frac{2\pi}{\lambda} x_{M-1}  u_x}
	\end{bmatrix}.
\end{equation}

According to \cite{RA-2ZBX}, the general antenna directional gain pattern is defined as
\begin{align}\label{antenna gain}
	G(\varphi)=
	\begin{cases}
		G_0 \cos^{2p}(\varphi), & \varphi \in \left[0, \frac{\pi}{2}\right) \\
		0, & \text{otherwise},
	\end{cases} 
\end{align}
where $\varphi$ is the boresight deflection angle between the emitter direction and the normal vector of the array, $G_0=2(2p+1)$ is the maximum gain in the normal direction (i.e., $\varphi$ = 0) that meets the law of power conservation. 

Considering that each element has the same antenna pattern and the free-space propagation path loss, the channel amplitude gain between the UAV and the BS is modeled as  
\begin{align}\label{g}
	g(\varphi)=
	\begin{cases}\sqrt{\frac{A}{4\pi r^2 }G(\varphi)}, & \varphi\in[0,\frac{\pi}{2}) \\
		0, & \text{otherwise},
	\end{cases} 
\end{align}
where the integral space $A$ corresponds to the surface region of rotatable array, and $r$ is the distance between emitter and array. 

To simplify, a constant is defined as $g_0 =\sqrt{\frac{A}{4\pi r^2 }G_0}$. Substituting Eq. (\ref{antenna gain}) into Eq. (\ref{g}), the effective channel gain is rewritten as
\begin{align}\label{g_simplified}
	g(\varphi)=
	\begin{cases}
		g_0 \cos^{p}(\varphi), & \varphi\in\left[0,\frac{\pi}{2}\right)\\
		0, & \text{otherwise}.
	\end{cases} 
\end{align}

Assume that the signal $s$, emitted by the UAV, is a narrowband signal, and $\mathbb{E}\{|s|^{2}\}=P_{t}$. Let $\mathbf{y}\in \mathbb{ C}^{ MN \times 1}$ be the received data vector as follows
\begin{equation}\label{y1}
	\mathbf{y}=s\,g(\varphi)\,\mathbf{a}(\theta,\phi) +\mathbf{n},
\end{equation}
where $\mathbf{n} \sim \mathcal{ CN}(0,\sigma^2\mathbf{I})$ is the additive white Gaussian noise (AWGN) vector.

Correspondingly, the received signal at the $(m,n)$-th antenna can be expressed as
\begin{align}\label{y}
	y_{m,n}&=sg(\varphi)e^{j\frac{2\pi}{\lambda}( x_m u_x+z_n u_z)} +n_{m,n},
\end{align}
where $\lambda$ is the wavelength and $n_{m,n}$ denotes the AWGN of the $(m,n)$-th antenna.

In Fig. \ref{fig:xyz}\subref{subfig:xyz-a}, the rotation array is placed on the $x–z$ plane. Therefore, the normal vector of array is aligned with the $y$-axis which is expressed as
\begin{align}
	\overrightarrow{OY_0}=[0,\,1,\,0]^T.
\end{align}

In accordance with the well-known cosine identity \cite{Gurney1961}, the boresight deflection angle $\varphi$ is
\begin{align}
	\varphi=\theta.
\end{align}

Now, let us consider the rotatable array model. 
The mechanical 3D rotation of the array is denoted by a rotation matrix $\mathbf{R} \in \mathbb{R}^{3 \times 3}$. Assuming the local coordinate system rotates synchronously with the array, the direction vector of the emitter in the local frame after rotation is given by
\begin{align}
	\mathbf{u}_{l}&=\mathbf{R}^T\mathbf{u}
	=[u_{l,x},u_{l,y},u_{l,z}]^T.
\end{align}

Based on this local coordinate system, the local boresight deflection angle becomes $\varphi_l = \theta_l = \arccos(u_{l,y})$. Consequently, the received signal at the $(m,n)$-th antenna after rotation is expressed as
\begin{align}\label{rotated signal}
	y_{l,m,n}&=sg(\theta_l)e^{j\frac{2\pi}{\lambda}( x_m u_{l,x}+z_n u_{l,z})} +n_{m,n}.
\end{align}

\section{Derived CRLB for Rotatable Array}
\noindent
To evaluate the performance of the proposed methods, the CRLB for the rotatable array is derived as a theoretical benchmark \cite{2009Classical, LC-CRLB}. For notational simplicity, the elevation and azimuth angles in the rotated coordinate system are simply denoted as $(\theta, \phi)$ in the following derivation.

Based on the received signal model in Eq. (\ref{rotated signal}), the observation at the $(m,n)$-th antenna element can be equivalently rewritten by separating the deterministic component and the noise, given by
\begin{equation}
	\tilde{y}_{m,n}=\tilde{\mu}_{m,n}+n_{m,n},
\end{equation}
where the deterministic signal part $\tilde{\mu}_{m,n}$ is defined as
\begin{equation}\label{mu}
	\tilde{\mu}_{m,n}=s\,g(\theta)\,e^{j\psi _{m,n}(\theta,\phi)},
\end{equation}
and the corresponding phase term is
\begin{equation}\label{psi}
	\begin{aligned}
		\psi _{m,n}(\theta,\phi)&=\frac{2\pi}{\lambda}( x_m u_x+z_nu_z)\\
		&=\frac{2\pi}{\lambda}( x_m \sin\theta \cos\phi+z_n\sin\theta\sin\phi).
	\end{aligned}
\end{equation}

The corresponding received signal vector of BS is
\begin{equation}
	\tilde{\mathbf{y}} = 
	\big[
	\tilde{y}_{0,0},\, \tilde{y}_{0,1},\dots,
	\tilde{y}_{M-1,N-1}
	\big]^{T}.
\end{equation}

Let us define the parameter vector as follows
\begin{equation}
	\boldsymbol{\theta}=[\theta,\phi]^{T}.
\end{equation}

Referring to \cite{2006Fundamentals}, the likelihood function of $\tilde{\mathbf{y}}$ is given by 
\begin{equation}
	p(\tilde{\mathbf{y}};\boldsymbol{\theta})
	=(\pi\sigma^{2})^{-MN}
	\exp\!\left(
	-\frac{\|\tilde{\mathbf{y}}
		-\boldsymbol{\mu}(\boldsymbol{\theta})\|^{2}}
	{\sigma^{2}}
	\right),
\end{equation}
where $\boldsymbol{\mu}(\boldsymbol{\theta})$ denotes the deterministic mean vector.

In order to derive the joint CRLB for the elevation and azimuth angles, the Fisher information matrix (FIM) \cite{1994Fundamentals} is defined as
\begin{equation}\label{FIM}	
	[\mathbf{F}(\boldsymbol{\theta })]_{ij}=-E\left [\frac{\partial ^2\ln p(\tilde{\mathbf{y}};\boldsymbol{\theta } )}{\partial\theta_i\partial\theta_j }\right ].
\end{equation}

Furthermore, Eq. (\ref{FIM}) is simplified as
\begin{align}	
	&[\mathbf{F}(\boldsymbol{\theta })]_{ij}=\frac{2}{\sigma^2}\sum_{m=0}^{M-1}\sum_{n=0}^{N-1}\Re\{\frac{\partial \tilde{\mu}_{m,n} }{\partial\theta_i}\cdot \frac{\partial \tilde{\mu}^*_{m,n} }{\partial\theta_j}\},
\end{align}
where 
\begin{align}
	&\Re\left\{ \left(\frac{\partial \tilde{\mu}_{m,n}}{\partial\boldsymbol{\theta}}\right) \left(\frac{\partial \tilde{\mu}_{m,n}}{\partial\boldsymbol{\theta}}\right)^H \right\}\\
	=&s^2\begin{bmatrix}
		\left(\frac{\partial g(\theta) }{\partial\theta}\right)^2+ g^2(\theta) \left(\frac{\partial \psi_{m,n} }{\partial\theta}\right)^2 & g^2(\theta) \frac{\partial \psi_{m,n}}{\partial \theta}\frac{\partial \psi_{m,n}}{\partial \phi}\\
		g^2(\theta) \frac{\partial \psi_{m,n}}{\partial \theta}\frac{\partial \psi_{m,n}}{\partial \phi} & g^2(\theta) \left(\frac{\partial \psi_{m,n} }{\partial\phi}\right)^2
	\end{bmatrix}.\nonumber
\end{align}

Differentiating $\psi_{m,n}$ with respect to $\theta$ and $\phi$ based on Eq. (\ref{psi}) yields
\begin{align}	
	\frac{\partial \psi_{m,n}}{\partial \theta} = k \cos\theta(x_m\cos\phi
	+ z_n\sin\phi),
\end{align}
and
\begin{align}	
	\frac{\partial \psi_{m,n}}{\partial \phi} = k\sin\theta(-x_m\sin\phi
	+ z_n\cos\phi),
\end{align}
where $k=\frac{2\pi}{\lambda} $.

Since $M$ and $N$ are odd numbers and the array elements are symmetrically distributed, we have $\sum_{m,n}x_m=0$, $\sum_{m,n}z_n=0$, $\sum_{m,n}x_m\,z_n=0$. Let us define
\begin{align}	
	S_m&=\sum^{M-1}_{m=0}x_m^2=\frac{M(M^2-1)}{12}d_x^2,\nonumber\\
	S_n&=\sum^{N-1}_{n=0}z_n^2=\frac{N(N^2-1)}{12}d_z^2.
\end{align}

Consequently, the single-snapshot FIM is derived as
\begin{align}
	\mathbf{F}_{\text{(1 snap)}}(\boldsymbol{\theta})
	&= \frac{2s^{2}}{\sigma^{2}}
	\begin{bmatrix}
		P & Q\\
		Q & R
	\end{bmatrix},
\end{align}
where
\begin{align}\label{PQR}	
	P &= MN\left(\frac{\partial g(\theta)}{\partial\theta}\right)^2 \nonumber\\&+ (kg(\theta)\cos\theta)^2\left[MS_n\sin^2\phi+NS_m\cos^2\phi \right], \nonumber \\
	Q &= (kg(\theta))^2\sin\theta\cos\theta\sin\phi\cos\phi\left[MS_n-NS_m \right],\nonumber \\
	R &= (kg(\theta)\sin\theta)^2\left[MS_n\cos^2\phi+NS_m\sin^2\phi \right].
\end{align}

Given $K$ independent measurements $\mathbf{y}[1]$,...,$\mathbf{y}[K]$, \textcolor{black}{where $K$ is the number of snapshots,} the total FIM is expressed as
\begin{align}\label{CRLB}
	\mathbf{F}(\boldsymbol{\theta})=\frac{2\,K\,s^2}{\sigma^2}\begin{bmatrix}
		P	& Q\\
		Q	& R
	\end{bmatrix}.
\end{align}

Differentiating Eq. (\ref{g_simplified}) with respect to $\theta$ yields
\begin{align}\label{g derivative theta}	
	\frac{\partial g(\theta)}{\partial\theta }=-g_0p \cos^{p-1}\theta\sin\theta .
\end{align}

By substituting Eq. (\ref{g derivative theta}) into Eq. (\ref{PQR}), the elements of the FIM are reformulated as
\begin{align}	
	P &= MN(g_0p \cos^{p-1}\theta\sin\theta)^2 \\&+ (kg_0\cos^{p+1}\theta)^2\left[MS_n\sin^2\phi+NS_m\cos^2\phi \right], \nonumber \\
	Q &= (kg_0\cos^p\theta)^2\sin\theta\cos\theta\sin\phi\cos\phi\left[MS_n-NS_m \right],\nonumber \\
	R &= (kg_0\cos^p\theta\sin\theta)^2\left[MS_n\cos^2\phi+NS_m\sin^2\phi \right].\nonumber
\end{align}

Therefore, the CRLB matrix is obtained by inverting the FIM, expressed as
\begin{equation}
	\text{CRLB} = \mathbf{F}^{-1}(\boldsymbol{\theta}) = \frac{\sigma^2}{2 K s^2 (PR - Q^2)} \begin{bmatrix}
		R & -Q\\
		-Q & P
	\end{bmatrix}.
\end{equation}
Let $\gamma=\frac{2Ks^2}{\sigma^2}$ denote the received SNR at the array.
The closed-form expressions of the theoretical lower bounds are subsequently derived as Eq. (\ref{CRLB of theta}) and Eq. (\ref{CRLB of phi}).
\begin{figure*} 
	\centering
	\begin{align}\label{CRLB of theta} 
		\text{CRLB}_{\theta}(\theta,\phi)
		&=
		\frac{\frac{1}{\gamma}
			(MS_n\cos^2\phi+NS_m\sin^2\phi)
		}{
			MN\,g_0^2
			\bigg[
			p^2\cos^{2p-2}\theta\,\sin^2\theta
			\Big(MS_n\cos^2\phi+NS_m\sin^2\phi\Big)
			+k^2\cos^{2p+2}\theta\,S_mS_n
			\bigg]
		}.\\
		\text{CRLB}_{\phi}(\theta,\phi)
		&=
		\frac{\frac{1}{\gamma}
			\Big(MN\,p^2\cos^{2p-2}\theta\,\sin^2\theta
			+
			k^2\cos^{2p+2}\theta
			(MS_n\sin^2\phi+NS_m\cos^2\phi)\Big)
		}{
			MN\,g_0^2k^2\cos^{2p}\theta\,\sin^2\theta
			\bigg[
			p^2\cos^{2p-2}\theta\,\sin^2\theta
			\Big(MS_n\cos^2\phi+NS_m\sin^2\phi\Big)
			+k^2\cos^{2p+2}\theta\,S_mS_n
			\bigg]
		}.\label{CRLB of phi}
	\end{align}
	\hrule 
\end{figure*}

As indicated by Eq. (\ref{CRLB of theta}) and Eq. (\ref{CRLB of phi}), the estimation performance is severely constrained by the directional antenna gain. When the \textcolor{black}{target} is aligned with the array boresight (i.e., $\theta \approx 0$), the antenna gain is maximized, yielding the minimum CRLB. Conversely, as the signal direction deviates from the normal vector, the effective SNR is significantly attenuated by the rapidly decreasing gain, which causes a sharp increase in the CRLB and a severe degradation in estimation accuracy. To overcome the performance loss outside the high-gain region, the array orientation must be mechanically adjusted to steer the target direction. This critical requirement directly motivates the pre-rotation initialization strategy proposed in the following section.

\section{Proposed Low-Complexity Rotatable  Array DOA Methods}\label{Section 4}

\begin{figure}
	\centering
	\includegraphics[width=0.9\linewidth]{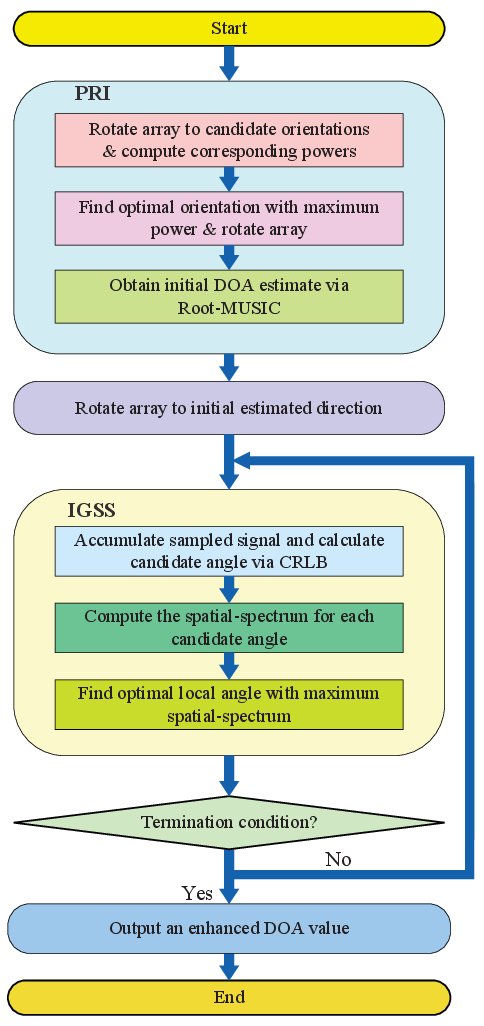}
	\caption{Flowchart of the proposed low-complexity enhanced direction-sensing method via a rotatable antenna array.}
	\label{Flowchart}
\end{figure}

In this section, to reduce the high complexity and enhance the estimation performance, a low-complexity DOA method is proposed. As illustrated in Fig. \ref{Flowchart}, the proposed framework consists of three main steps: PRI is proposed in first subsection, rotating the array to the initial estimate direction, and IGSS is proposed in second subsection. Finally, a detailed computational complexity analysis is presented.

\subsection{Proposed Low-Complexity PRI Method}

As the emitter deviates from the array boresight, the array gain decreases rapidly due to the directive antenna pattern, causing a severe attenuation in the received SNR and a consequent degradation in estimation accuracy. To overcome this limitation, a low-complexity PRI method is introduced in this subsection. Specifically, the array orientation is coarsely adjusted to an optimal direction with the maximum received power, thereby significantly enhancing the effective SNR.

Specifically, the array is rotated around the $z$-axis and the $x$-axis. The discrete candidate scanning angle sets for these rotations are defined as
\begin{align} 
	\Psi_{\alpha} &= \{ \alpha_1, \dotsc, \alpha_Q \},\\
	\Psi_{\beta} &= \{ \beta_1, \dotsc, \beta_Q \}, 
\end{align}
where $\Psi_{\alpha}$ and $\Psi_{\beta}$ correspond to the rotation angles around the $z$-axis and the $x$-axis, respectively, and $Q$ denotes the number of uniformly distributed angles in each dimension.

For a specific pair of candidate angles $(\alpha_i, \beta_j)$, the elementary rotation matrices are given by
\begin{align} \label{R of a}
	\mathbf{R}_z(\alpha_i) &= \begin{bmatrix} \cos\alpha_i & -\sin\alpha_i & 0 \\ \sin\alpha_i & \cos\alpha_i & 0 \\ 0 & 0 & 1 \end{bmatrix}, \\
	\mathbf{R}_x(\beta_j) &= \begin{bmatrix} 1 & 0 & 0 \\ 0 & \cos\beta_j & -\sin\beta_j \\ 0 & \sin\beta_j & \cos\beta_j \end{bmatrix}. \label{R of b}
\end{align}

The combined 3D rotation matrix is expressed as
\begin{equation} \label{R of ab}
	\mathbf{R}(\alpha_i, \beta_j) = \mathbf{R}_z(\alpha_i) \mathbf{R}_x(\beta_j).
\end{equation}

By traversing all combinations of candidate directions, a pre-rotation candidate set comprising $Q^2$ orientations is constructed as
\begin{equation}
	\Omega = \left\{ \mathbf{R}_q \mid \mathbf{R}_q = \mathbf{R}(\alpha_i, \beta_j), \forall \alpha_i \in \Psi_{\alpha}, \beta_j \in \Psi_{\beta} \right\},
\end{equation}
where $q=1,2,\dots,Q^2$ is the index of the candidate orientations.

Since $Q$ is deliberately set to a small integer for coarse initialization, the size of the search space is significantly constrained, which theoretically guarantees the low computational complexity of the PRI stage.

\begin{figure}
	\centering
	\includegraphics[width=1\linewidth]{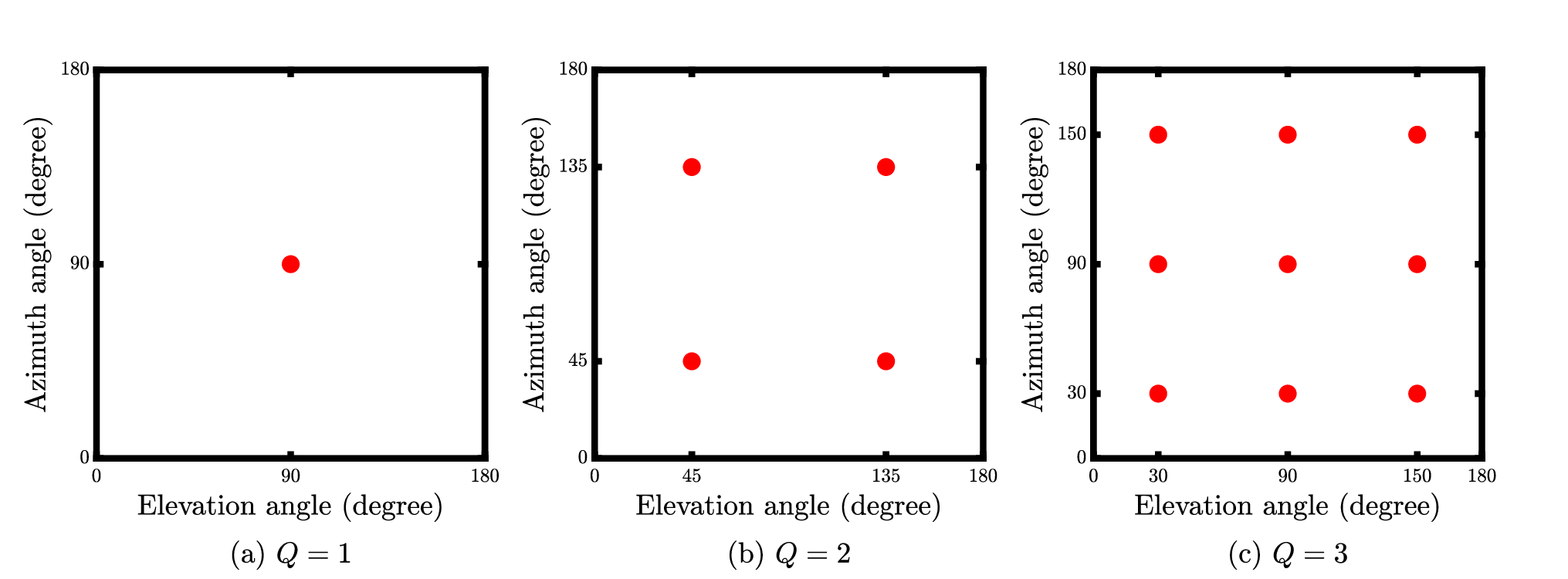}
	\caption{Illustration of the spatial distribution of the different pre-rotation candidate set $\Omega$.}
	\label{fig:pre-rotation-diagram}
\end{figure}

To demonstrate the pre-rotation process, Fig.  \ref{fig:pre-rotation-diagram} illustrates the spatial distribution of the pre-rotation candidate set $\Omega$. It can be observed that as $Q$ increases, the points within the candidate set $\Omega$ become denser, thereby providing finer coverage of high-gain search directions within the given fixed angular range.

For each candidate orientation $\mathbf{R}_q \in \Omega$, the received signal is collected over $K$ snapshots. The corresponding received power is computed as
\begin{align}
	P(\mathbf{R}_q) = \sum_{k=1}^{K} \left\| \mathbf{y
	}_q[k] \right\|_2^2,
\end{align}
where $\mathbf{y}_q[k]$ denotes the $k$-th snapshot of the received signal vector defined in Eq. (\ref{y1}) under the $q$-th orientation. 

The optimal rotation matrix, denoted as $\mathbf{R}^\star$, is determined by maximizing the received power as follows
\begin{equation}
	\mathbf{R}^\star = \arg\max_{\mathbf{R}_q \in \Omega} P(\mathbf{R}_q).
\end{equation}

Subsequently, the array is mechanically rotated to this optimal direction. Based on the optimally rotated array, the sample covariance matrix of the received signal is calculated as
\begin{equation}\label{R}
	\hat{\mathbf{R}}_{yy} = \frac{1}{K}\sum_{k=1}^{K} \mathbf{y}[k]\mathbf{y}^H[k],
\end{equation}
where $\mathbf{y}[k]$ is the received signal at the $k$-th snapshot.

Finally, the EVD of $\hat{\mathbf{R}}_{yy}$ is performed as
\begin{equation}	
	\hat{\mathbf{R}}_{yy} =[\hat{\mathbf{U}}_{s}, \hat{\mathbf{U}}_{n}] \mathbf{\Sigma} [\hat{\mathbf{U}}_{s}, \hat{\mathbf{U}}_{n}]^H,
\end{equation}
where $\mathbf{\Sigma}$ is the diagonal matrix of eigenvalues, $\hat{\mathbf{U}}_{s} = [\hat{\mathbf{u}}_1] \in \mathbb{C}^{MN \times 1}$ is the estimated signal subspace, and $\hat{\mathbf{U}}_{n} = [\hat{\mathbf{u}}_2, \dots, \hat{\mathbf{u}}_{MN}] \in \mathbb{C}^{MN \times (MN-1)}$ is the estimated noise subspace. 

Based on the MUSIC principle, the array manifold vector is orthogonal to the noise subspace. Thus, the spatial cost function of the 2D-MUSIC algorithm is formulated as
\begin{equation}\label{Orthogonality condition}	
	\mathcal{J}(\boldsymbol{\theta}) = \sum_{i=2}^{MN} \left| \hat{\mathbf{u}}_i^H \left( \mathbf{a}_z(u_z) \otimes \mathbf{a}_x(u_x) \right) \right|^2.
\end{equation}

To avoid an exhaustive 2D search and reduce computational complexity, the noise eigenvector is reshaped into a matrix $\hat{\mathbf{N}}_i \in \mathbb{C}^{M \times N}$ such that $\hat{\mathbf{u}}_i = \text{vec}(\hat{\mathbf{N}}_i)$. Consequently, the cost function in Eq. (\ref{Orthogonality condition}) is rewritten as
\begin{equation}\label{Orthogonality}
	\begin{aligned}
		\mathcal{J}(\boldsymbol{\theta}) &= \sum_{i=2}^{MN} \left| \mathbf{a}_x^H(u_x) \hat{\mathbf{N}}_i^H \mathbf{a}_z(u_z) \right|^2 \\
		&= \mathbf{a}_x^H(u_x) \left( \sum_{i=2}^{MN} \hat{\mathbf{N}}_i \big(\mathbf{a}_z(u_z)\mathbf{a}_z^H(u_z)\big) \hat{\mathbf{N}}_i^H \right) \mathbf{a}_x(u_x).
	\end{aligned}
\end{equation}

By substituting the approximation $\mathbf{a}_z(u_z)\mathbf{a}_z^H(u_z) \approx \mathbf{I}_N$ into (\ref{Orthogonality}), the original 2D coupled problem is decoupled into two independent 1D Root-MUSIC problems. The decoupled cost function for the $x$-dimension is given by
\begin{equation}\label{J_ux}	
	\mathcal{J}_x(u_x) = \mathbf{a}_x^H(u_x) \left( \sum_{i=2}^{MN} \hat{\mathbf{N}}_i \hat{\mathbf{N}}_i^H \right) \mathbf{a}_x(u_x).
\end{equation}

Similarly, the cost function for the $z$-dimension is decoupled as
\begin{equation}\label{J_uz}		
	\mathcal{J}_z(u_z) = \mathbf{a}_z^H(u_z) \left( \sum_{i=2}^{MN} \hat{\mathbf{N}}_i^H \hat{\mathbf{N}}_i \right) \mathbf{a}_z(u_z).
\end{equation}

In accordance with Eq. (\ref{J_ux}) and Eq. (\ref{J_uz}), the local direction cosines $\hat{u}_x$ and $\hat{u}_z$ are independently estimated. Subsequently, the local elevation and azimuth angles are reconstructed as
\begin{align}
	\label{local_theta} \hat{\theta}_l &= \arccos\left(\sqrt{1 - \hat{u}_z^2 - \hat{u}_x^2}\right), \\
	\label{local_phi} \hat{\phi}_l &= \operatorname{atan2}(\hat{u}_z, \hat{u}_x).
\end{align}

Let $\hat{\mathbf{u}}_l$ denote the local direction vector constructed from these local angles. The initial global direction vector is recovered by mapping it back to the global coordinate system via the previously selected optimal rotation matrix $\mathbf{R}^\star$:
\begin{equation}
	\hat{\mathbf{u}}_{\text{init}} = \mathbf{R}^\star \hat{\mathbf{u}}_l.
\end{equation}

Finally, the initial global DOA estimate $\hat{\boldsymbol{\theta}}_{\text{init}}$ is extracted from the elements of $\hat{\mathbf{u}}_{\text{init}}$ using the inverse trigonometric mapping defined in Eq. (\ref{local_theta}) and Eq. (\ref{local_phi}).

\subsection{Proposed Low-Complexity IGSS Method}

In this stage, an IGSS method is introduced to further enhance the estimation accuracy through an iterative greedy search and continuous snapshot accumulation.

Based on the initial global direction estimate $\hat{\mathbf{u}}_{\text{init}} = [\hat{u}_{\text{init},x}, \hat{u}_{\text{init},y}, \hat{u}_{\text{init},z}]^T$, the mechanical rotation angles required to align the array boresight with the target are computed as
\begin{align}
	\alpha_{\text{init}} &= \operatorname{atan2}(-\hat{u}_{\text{init},x}, \hat{u}_{\text{init},y}), \\
	\beta_{\text{init}} &= \arcsin(\hat{u}_{\text{init},z}).
\end{align}

Then the corresponding rotation matrix is given by
\begin{equation}
	\mathbf{R}_{\text{init}} = \mathbf{R}_z(\alpha_{\text{init}}) \mathbf{R}_x(\beta_{\text{init}}).
\end{equation}

The array is mechanically rotated to the initial direction using $\mathbf{R}_{\text{init}}$ and keeps fixed for the remainder of the estimation process.

Since the array normal vector has been aligned with the initial global estimate $\hat{\mathbf{u}}_{\text{init}}$ via mechanical rotation, the initial local DOA estimate in the rotated coordinate system is naturally set as the origin, i.e.,
\begin{equation}
	\hat{\boldsymbol{\theta}}_{l, 0} = [0, 0]^T.
\end{equation}

This initialization serves as the starting point for the subsequent greedy search process.

Under this fixed orientation, the spatial spectrum is evaluated in the local coordinate system. For any candidate local direction $\boldsymbol{\theta}_l = [\theta_l, \phi_l]^T$, the local direction vector is denoted as $\mathbf{u}_l(\theta_l, \phi_l)$. Consequently, the corresponding local steering vector $\mathbf{a}(\boldsymbol{\theta}_l)$ is directly constructed as
\begin{equation} 
	\mathbf{a}(\boldsymbol{\theta}_l) = \left[ \mathrm{e}^{\mathrm{j} \frac{2\pi}{\lambda} \mathbf{p}_{0,0}^T \mathbf{u}_{l}}, \dots, \mathrm{e}^{\mathrm{j} \frac{2\pi}{\lambda} \mathbf{p}_{M-1,N-1}^T \mathbf{u}_{l}} \right]^T. 
\end{equation}

Assuming the array continuously collects snapshots. Let $\mathbf{Y}_i\in \mathbb{C} ^{MN\times K}$ denote the new snapshot matrix collected during the $i$-th iteration stage. The accumulated data matrix up to $i$-th iteration is defined as
\begin{align} 
	\mathbf{Y}^{(i)}=[\mathbf{Y}_1,...,\mathbf{Y}_i]\in \mathbb{C} ^{MN\times iK}.
\end{align}

The corresponding sample covariance matrix is calculated by
\begin{equation}	
	\hat{\mathbf{R}}_{yy}^{(i)} = \frac{1}{iK} \mathbf{Y}^{(i)} \left(\mathbf{Y}^{(i)}\right)^H.
\end{equation}

To circumvent the computational burden of directly computing the covariance matrix from scratch, $\hat{\mathbf{R}}_{yy}^{(i)}$ is equivalently updated in a recursive manner as follows
\begin{equation}	
	\hat{\mathbf{R}}_{yy}^{(i)} = \frac{i-1}{i} \hat{\mathbf{R}}_{yy}^{(i-1)} + \frac{1}{iK} \mathbf{Y}_i \mathbf{Y}_i^H.
\end{equation}

As the number of effective snapshots increases from $K$ to $iK$, the FIM grows proportionally, leading to a monotonically decreasing CRLB.

To balance convergence speed and precision, a variable step-size strategy guided by the closed-form CRLB is adopted. Let $\hat{\boldsymbol{\theta}}_{l, i-1}$ denote the estimated local DOA from the $(i-1)$-th iteration. The local search step sizes $\Delta \boldsymbol{\theta}_i = [\Delta_{\theta, i}, \Delta_{\phi, i}]^T$ for the $i$-th iteration are designed as
\begin{align}
	\Delta_{\theta, i} &= \min \left( \sqrt{\mathrm{CRLB}_{\theta}(\hat{\boldsymbol{\theta}}_{l, i-1}; iK)} \cdot \gamma^{r_i}, \Delta_{\max} \right), \\
	\Delta_{\phi, i} &= \min \left( \sqrt{\mathrm{CRLB}_{\phi}(\hat{\boldsymbol{\theta}}_{l, i-1}; iK)} \cdot \gamma^{r_i}, \Delta_{\max} \right),
\end{align}
where $\gamma \in (0, 1)$ is a decay factor, and $r_i$ is a refinement index. The CRLB is evaluated in the local array coordinate system.

The candidate set $\boldsymbol{\Omega}_{l,i}$ for the current iteration is defined around the previous local estimate as
\begin{equation}
	\boldsymbol{\Omega}_{l, i} = \left\{ \hat{\boldsymbol{\theta}}_{l, i-1} + \mathbf{d} \odot \Delta\boldsymbol{\theta}_{i} \mid \mathbf{d} \in \mathcal{D} \right\},
\end{equation}
where $\mathcal{D}$ denotes the directional search space as follows
\begin{align}
	\mathcal{D} = \left\{ 
	\begin{bmatrix}0 \\ 0\end{bmatrix}, 
	\pm \begin{bmatrix}1 \\ 0\end{bmatrix}, 
	\pm \begin{bmatrix}0 \\ 1\end{bmatrix}
	\right\}.
\end{align}

Subsequently, the updated local estimate is obtained by maximizing the spatial spectrum as follows
\begin{equation}\label{IGSS}
	\hat{\boldsymbol{\theta}}_{l, i} = \arg\max_{\boldsymbol{\theta}_l \in \boldsymbol{\Omega}_{l, i}} \mathbf{a}^H(\boldsymbol{\theta}_l) \hat{\mathbf{R}}_{yy}^{(i)} \mathbf{a}(\boldsymbol{\theta}_l).
\end{equation}

Based on the newly obtained estimate, the search refinement index for the next iteration is updated as
\begin{equation} 
	r_{i+1} = \begin{cases} r_{i} + 1, & \text{if } \boldsymbol{\theta}_{l,i} = \boldsymbol{\theta}_{l,i-1}, \\ 
		r_{i}, & \text{otherwise}, \end{cases}
\end{equation} 
where the initial index is set to $r_1 = 0$, and $r_i$ is upper-bounded by $r_{\max}$.

The iterative search process terminates when the sequence of local DOA estimates converges, which is evaluated by
\begin{equation}
	\left\| \hat{\boldsymbol{\theta}}_{l, i} - \hat{\boldsymbol{\theta}}_{l, i-1} \right\|_2 \le \epsilon,
\end{equation}
where $\epsilon$ is a predefined small tolerance threshold. To prevent infinite loops in cases of severe noise or search stagnation, the algorithm is also forced to halt if the number of iterations reaches a maximum value $I_{\max}$ or the refinement index exceeds its upper bound, i.e., $r_i > r_{\max}$.

Let $\hat{\boldsymbol{\theta}}_{l, \text{final}} = [\hat{\theta}_{l, \text{final}}, \hat{\phi}_{l, \text{final}}]^T$ denote the estimated local DOA upon convergence. The corresponding local direction vector is constructed and denoted as $\hat{\mathbf{u}}_{l, \text{final}}$. 

To obtain the final DOA estimate in the global coordinate system, the local direction vector is mapped back using the previously fixed mechanical rotation matrix $\mathbf{R}_{\text{init}}$ as follows
\begin{equation}
	\hat{\mathbf{u}}_{\text{final}} = \mathbf{R}_{\text{init}} \hat{\mathbf{u}}_{l, \text{final}}.
\end{equation}

Finally, the ultimate global DOA estimate, denoted as $\hat{\boldsymbol{\theta}}_{\text{final}} = [\hat{\theta}_{\text{final}}, \hat{\phi}_{\text{final}}]^T$, is extracted from the elements of $\hat{\mathbf{u}}_{\text{final}}$ using the same inverse trigonometric mapping defined in Eq. (\ref{local_theta}) and Eq. (\ref{local_phi}).

\subsection{Computational Complexity Analysis}
Now, the approximate computational complexities of the three rotatable array methods are analyzed. 
\textcolor{black}{Let $L=MN$ denote the total number of antenna elements in the UPA, which is also the dimension of the received signal vector. Thus, the sample covariance matrix has size $L\times L$. Moreover, $K$ is the number of snapshots, $Q$ is the number of candidate pre-rotation angles, and $I$ is the number of rotation/refinement iterations. In the IGSS stage, $J=|\mathcal{D}|$ denotes the number of local candidate directions evaluated in each iteration.} First, the complexity of the traditional RR-Root-MUSIC method is about $C_{RR-Root-MUSIC} = \mathcal{O}\big(I (K L^2 + L^3)\big)$ floating-point operations (FLOPs). For the proposed PRI, its computational complexity is approximated as $C_{PRI} = \mathcal{O}\big(Q^2 K L + K L^2 + L^3\big)$. For the proposed PRI-IGSS, the iterative greedy search is executed upon the initialization of PRI. Accordingly, the complexity of the proposed PRI-IGSS is approximated as $C_{PRI-IGSS} = C_{PRI} + \mathcal{O}\big(I(K L^2 + J L^2)\big)= \mathcal{O}\big(Q^2 K L + K L^2 + L^3+I(K L^2 + J L^2)\big)$, which is slightly higher than that of proposed PRI method. Generally, their computational complexities are in decreasing order as follows: RR-Root-MUSIC, PRI-IGSS, and PRI.

\section{Simulation Results}
In this section, we present simulation results to evaluate the performance of the proposed PRI and PRI-IGSS methods, \textcolor{black}{with the corresponding CRLB used as a performance benchmark}. \textcolor{black} {FA-Root-MUSIC denotes the conventional fixed-array Root-MUSIC method without mechanical rotation, which is used to evaluate the effect of off-boresight gain loss. RR-Root-MUSIC follows the recursive-rotation Root-MUSIC method in [34], where Root-MUSIC estimation and array rotation are alternately performed until convergence, and is used as a rotatable-array-aided benchmark.   
	The proposed PRI denotes the pre-rotation initialization followed by Root-MUSIC estimation without IGSS refinement, while PRI-IGSS denotes the complete proposed method.} Simulation parameters are chosen as follows: $N=M=7$, $P_t$= 5 dBm, $\sigma^2 $= 10 dBm, $p=2$, $\phi=45^\circ$, $\lambda$ = 0.125, $K$ = 512, $Q=3$, $\Psi_{\alpha} = \Psi_{\beta} = \{ 30^\circ, 90^\circ, 150^\circ \}$, \textcolor{black}{$\epsilon=10^{-5}$, $I_{\max}=30$, $\gamma=0.5$, and $\Delta_{\max}=1^\circ$}. In our simulation, all results are averaged over 2000 Monte Carlo realizations and root mean squared error (RMSE) is calculated to indicate the performance, which is given by
\begin{align}
	\text{RMSE} = \sqrt{\frac{1}{U}\sum_{u}^{U}(\theta_u-\theta)^2},
\end{align}	
where $U$ denotes the number of Monte Carlo experiments. \textcolor{black}{ It is assumed that the UAV is in a quasi-static state during the snapshot collection period. }

\begin{figure}[h]
	\centering
	\includegraphics[width=1\linewidth]{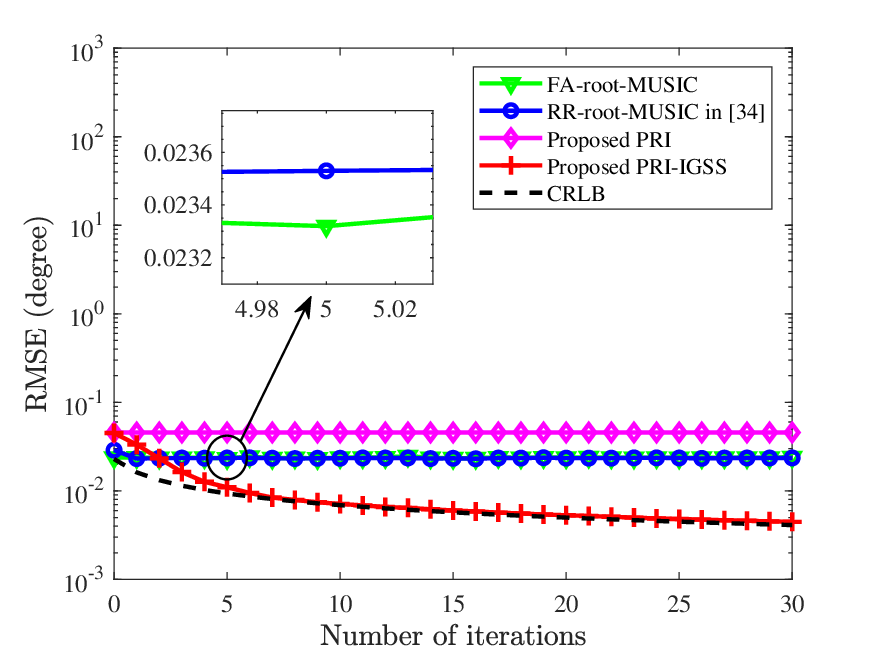}
	\caption{Convergence curves of the proposed low-complexity methods for rotatable array when $\theta = 5^o$.}
	\label{fig:iter theta=5}
\end{figure}

\begin{figure}
	\centering
	\includegraphics[width=1\linewidth]{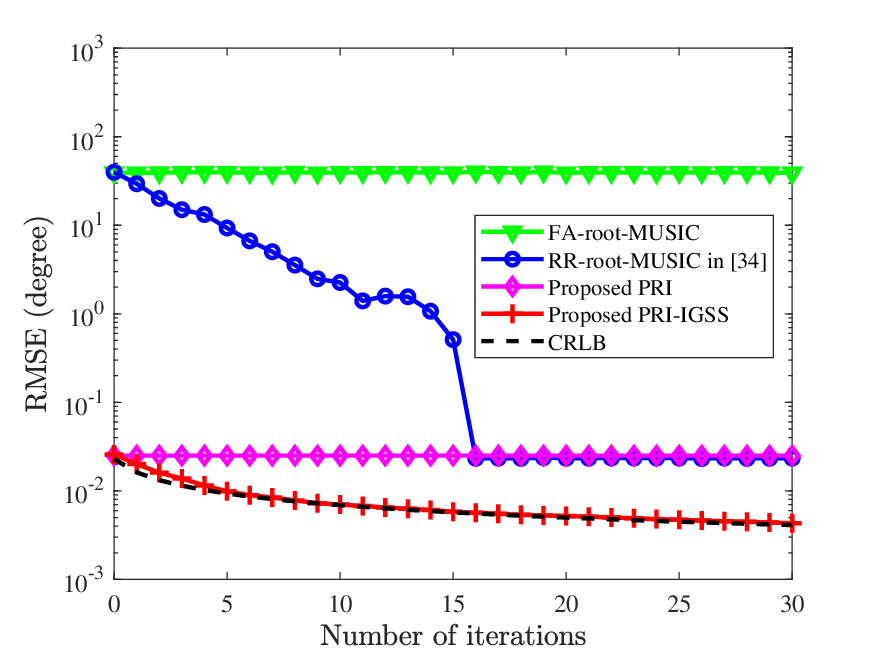}
	\caption{Convergence curves of the proposed low-complexity methods for rotatable array when $\theta = 85^o$.}
	\label{fig:iter theta=85}
\end{figure}
\begin{figure}[h]
	\centering
	\includegraphics[width=1\linewidth]{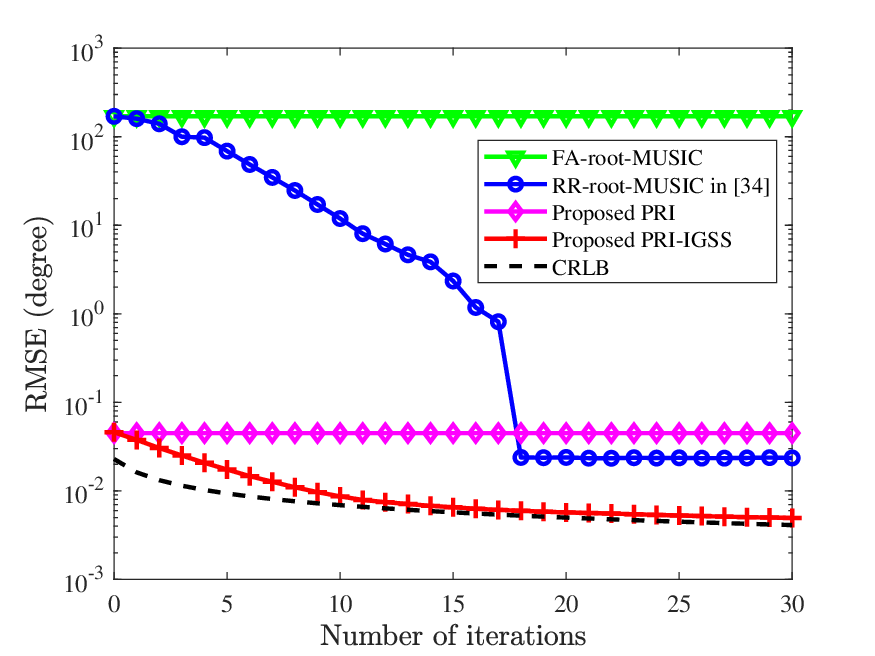}
	\caption{Convergence curves of the proposed low-complexity methods for rotatable array when $\theta = 175^o$.}
	\label{fig:iter theta=175}
\end{figure}

Fig. \ref{fig:iter theta=5}-\ref{fig:iter theta=175} illustrate the RMSE convergence curves of the proposed low-complexity PRI and PRI-IGSS methods for three representative incident angles, i.e., $\theta = 5^\circ, 85^\circ, 175^\circ$. 
It is seen that the CRLB \textcolor{black}{decreases as iteration process}  due to the  accumulation of sampling signals. \textcolor{black}{In Fig.5, where} $\theta = 5^\circ$, the conventional FA-Root-MUSIC method is completely disabled at large angles, i.e., $\theta = 85^\circ, 175^\circ$, due to severe SNR \textcolor{black}{attenuation induced by the directive property  of array element}. Although the RR-Root-MUSIC method eventually converges, it requires multiple rotation iterations for larger boresight deflection angle ($\theta = 85^\circ, 175^\circ$), resulting in high computational complexity, and still has a \textcolor{black}{large} gap compared to the CRLB. By contrast, the proposed PRI method effectively improves the initialization accuracy by pre-rotation and Root-MUSIC algorithm. On this basis, the PRI-IGSS method further improves the estimation through iterative greedy search with a continuous sampled signal accumulation process, leading to a rapid convergence and \textcolor{black}{an RMSE performance approaching the CRLB.}

\begin{figure}[h]
	\centering
	\includegraphics[width=1\linewidth]{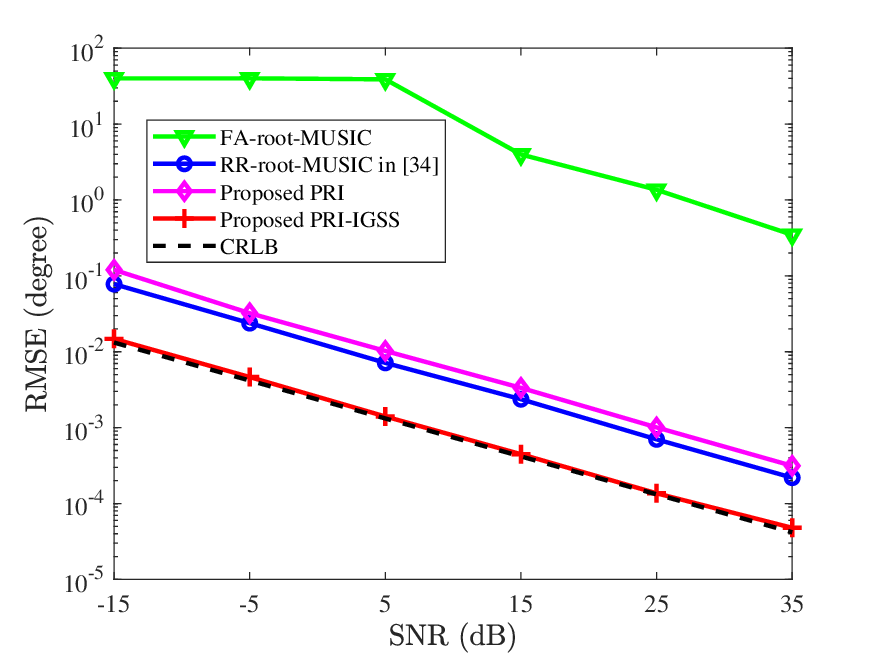}
	\caption{RMSE versus SNR of the proposed methods when $\theta=85^o$.}
	\label{fig:SNR}
\end{figure}

\begin{figure}[h]
	\centering
	\includegraphics[width=1\linewidth]{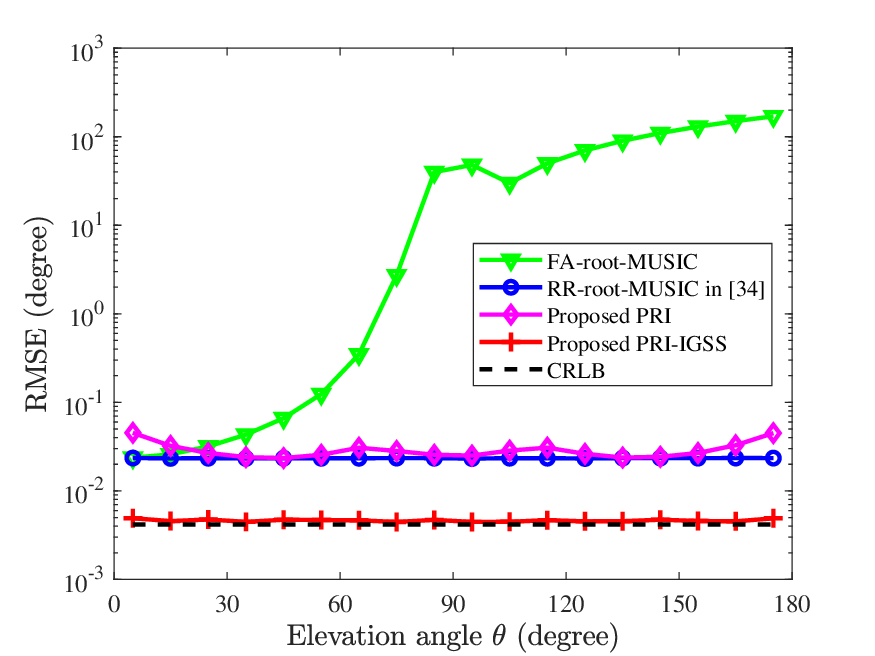}
	\caption{RMSE versus the elevation angle $\theta$ for the proposed methods when SNR = -5 dB.}
	\label{fig:theta}
\end{figure}

Fig. \ref{fig:SNR} shows the performance curves of RMSE versus SNR of the proposed methods with $\theta=85^o$.  \textcolor{black}{Due to the severe gain attenuation of the directional antenna pattern, the performance of the FA-Root-MUSIC remains almost constant in the low SNR region from $-15$ dB to $5$ dB. As the SNR further increases, the effective received SNR of FA-Root-MUSIC gradually becomes sufficient, and its RMSE decreases. The proposed PRI and the traditional RR-Root-MUSIC can provide stable estimation performance, but a noticeable gap from the CRLB still remains.}  In contrast, the CRLB is perfectly achieved by the proposed PRI-IGSS across the entire SNR range from -15 dB to 35 dB. Consequently, \textcolor{black}{nearly one-order-of-magnitude} improvement in sensing accuracy is successfully achieved by the proposed PRI-IGSS over the PRI and RR-Root-MUSIC.

\begin{figure}
	\centering
	\includegraphics[width=1\linewidth]{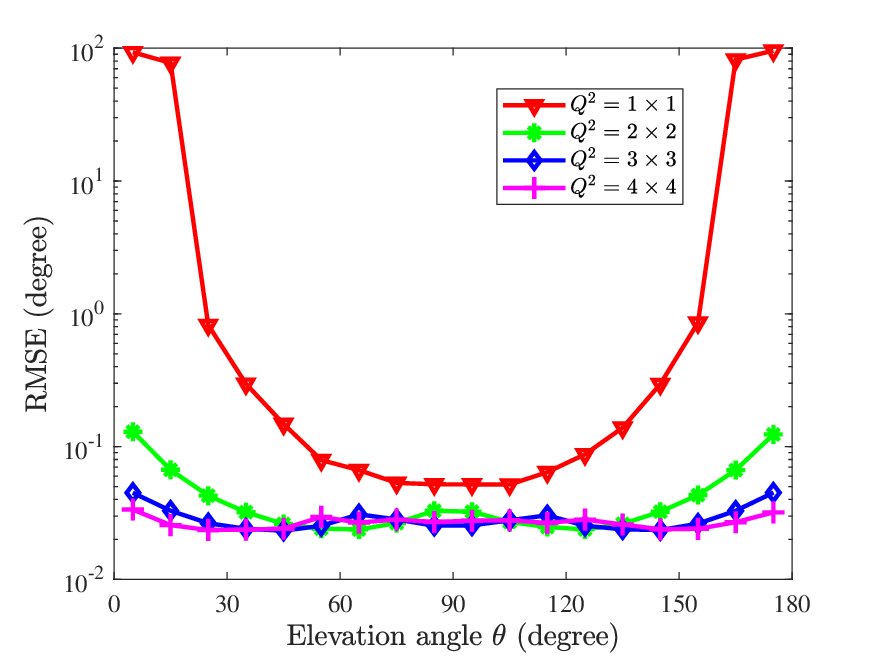}
	\caption{RMSE versus the elevation angle $\theta$ for the different number of candidate pre-rotation direction with SNR = 5 dB.}
	\label{fig:pre-rotation}
\end{figure}

In Fig. \ref{fig:theta}, the RMSE performance versus the elevation angle $\theta$ at a low SNR of -5 dB is evaluated. 
As $\theta$ increases or the target deviates from the array boresight, a severe performance degradation is suffered by the FA-Root-MUSIC, particularly in off-boresight regions, i.e., $\theta>80^o$. \textcolor{black}{In contrast, the proposed PRI and PRI-IGSS methods can effectively mitigate this degradation by rotating the array boresight toward the UAV direction. Compared with FA-Root-MUSIC, PRI and PRI-IGSS achieve up to about three-order and four-order magnitude improvements in RMSE, respectively.} Moreover, over the entire angular range, the PRI-IGSS consistently outperforms the RR-Root-MUSIC by about one-order magnitude. These results show that PRI-IGSS still has excellent spatial stability and excellent estimation performance even in extreme cases.

%\begin{figure}
%	\centering
%	\includegraphics[width=1\linewidth]{figures/Pre_rotation_SNR}
%	\caption{}
%	\label{fig:prerotationsnr}
%\end{figure}

Fig. \ref{fig:pre-rotation} plots the RMSE performance versus the elevation angle $\theta$ for the different number of candidate pre-rotation directions at a SNR of -5 dB. Based on equi-spaced rotation or sampling, the candidate angle sets for $Q=1, 2, 3,$ and $4$ are respectively defined as $\{90^\circ\}$, $\{45^\circ,135^\circ\}$, $\{30^\circ,90^\circ,150^\circ\}$, and $\{22.5^\circ,67.5^\circ,112.5^\circ,157.5^\circ\}$. As illustrated, a negligible performance difference between $Q=3$ and $Q=4$ is observed. This implies that the main high-gain region of the array is sufficiently covered by the candidate set of $Q=3$. Consequently, only trivial accuracy improvements are reaped by further increasing $Q$, while the number of rotation is increased, i.e., from 9 to 16. Thus, $Q=3$ \textcolor{black}{may be chosen as a good value} to balance sensing precision and computational burden.

\section{Conclusions}
In this paper, to address the critical challenges of high computational complexity and low time efficiency in practical rotatable array systems, we proposed a low-complexity enhanced direction-sensing framework with a pre-rotation operation, which will provide a good initial value for the following  iterative sensing step and will speed up the convergent process. Furthermore, the corresponding CRLB was derived to provide a rigorous theoretical benchmark for the rotatable array model. The proposed framework is divided into three stages: first, a finite number of equal-space samplings are utilized to obtain a highly reliable initial estimate in the PRI stage. Subsequently, the array is rotated to the initial estimate direction. Finally, in the IGSS stage, a receive beamforming is performed on the candidate angles to find the maximum spatial spectrum, and by adopting a gradually decreasing search step size, the estimation accuracy is continuously improved. Crucially, the proposed PRI-IGSS method dramatically decreases the repeated use of EVD of  Root-MUSiC  to sensing the direction, which reduces the computational complexity.  Through simulation and analysis, we found that the proposed PRI-IGSS can achieve the CRLB. Compared to the traditional FA-Root-MUSIC method, it improved the performance by up to four orders of magnitude at extreme elevation angles, for example, $\theta > 80^\circ$. Moreover, by exploiting continuous signal sampling accumulation, the proposed method exceeds the RR-Root-MUSIC approach by approximately one-order magnitude. Due to its merits in low complexity and superior spatial accuracy, the PRI-IGSS method will be very suitable for the future Low-Altitude Wireless Network to implement a high-precision UAV direction sensing and perform a directive beam towards UAVs with an enhanced high energy efficiency.

\bibliographystyle{IEEEtran}
\bibliography{refs}

% Generated by IEEEtran.bst, version: 1.12 (2007/01/11)
\begin{thebibliography}{10}
\providecommand{\url}[1]{#1}
\csname url@samestyle\endcsname
\providecommand{\newblock}{\relax}
\providecommand{\bibinfo}[2]{#2}
\providecommand{\BIBentrySTDinterwordspacing}{\spaceskip=0pt\relax}
\providecommand{\BIBentryALTinterwordstretchfactor}{4}
\providecommand{\BIBentryALTinterwordspacing}{\spaceskip=\fontdimen2\font plus
\BIBentryALTinterwordstretchfactor\fontdimen3\font minus
  \fontdimen4\font\relax}
\providecommand{\BIBforeignlanguage}[2]{{%
\expandafter\ifx\csname l@#1\endcsname\relax
\typeout{** WARNING: IEEEtran.bst: No hyphenation pattern has been}%
\typeout{** loaded for the language `#1'. Using the pattern for}%
\typeout{** the default language instead.}%
\else
\language=\csname l@#1\endcsname
\fi
#2}}
\providecommand{\BIBdecl}{\relax}
\BIBdecl

\bibitem{Low-Altitude-apply1}
G.~Cheng, X.~Song, Z.~Lyu, and J.~Xu, ``Networked isac for low-altitude
  economy: Coordinated transmit beamforming and uav trajectory design,''
  \emph{IEEE Transactions on Communications}, vol.~73, no.~8, pp. 5832--5847,
  2025.

\bibitem{Low-Altitude-apply2}
H.~Yuhong, D.~Haiyu, C.~Weiyan, K.~Luting, D.~Wei, L.~Xin, L.~Yang, W.~Guizhen,
  and L.~Liang, ``Towards a low-altitude aerial intelligent network: Vision,
  challenges, and key technologies,'' \emph{China Communications}, vol.~22,
  no.~9, pp. 1--21, 2025.

\bibitem{Low-Altitude}
S.~Gao, J.~Yan, P.~Huang, Z.~Lu, M.~Gong, L.~Miao, G.~Zhu, J.~Liang, and
  L.~Yang, ``Integrated sensing, communication, and computation for
  low-altitude networks towards seamless connectivity and connected
  intelligence,'' \emph{IEEE Internet of Things Magazine}, pp. 1--9, 2026.

\bibitem{Low-Altitude-Sensing-demand}
X.~Xia, Z.~Fang, K.~Xu, and W.~Xie, ``An ambiguity-function-assisted active
  sensing scheme for ofdm-based isac systems toward low-altitude airspace,''
  \emph{IEEE Internet of Things Journal}, vol.~12, no.~12, pp.
  19\,471--19\,487, 2025.

\bibitem{UAV-ISAC-YY}
Y.~Yao, W.~Xiao, P.~Miao, G.~Chen, H.~Yang, C.-B. Chae, and K.-K. Wong,
  ``Uav-rhs-enabled full-duplex isac covert system: Robust beamforming and
  trajectory optimization,'' \emph{IEEE Transactions on Communications},
  vol.~74, pp. 5637--5653, 2026.

\bibitem{UAV-XZY}
Z.~Xiao, L.~Zhu, and X.-G. Xia, ``Uav communications with millimeter-wave
  beamforming: Potentials, scenarios, and challenges,'' \emph{China
  Communications}, vol.~17, no.~9, pp. 147--166, 2020.

\bibitem{LC-DOA-SF}
F.~Shu, Y.~Qin, T.~Liu, L.~Gui, Y.~Zhang, J.~Li, and Z.~Han, ``Low-complexity
  and high-resolution doa estimation for hybrid analog and digital massive mimo
  receive array,'' \emph{IEEE Transactions on Communications}, vol.~66, no.~6,
  pp. 2487--2501, 2018.

\bibitem{HAD-SF}
F.~Shu, B.~Shi, Y.~Chen, J.~Bai, Y.~Li, T.~Liu, Z.~Han, and X.~You, ``A new
  heterogeneous hybrid massive mimo receiver with an intrinsic ability of
  removing phase ambiguity of doa estimation via machine learning,'' \emph{IEEE
  Transactions on Machine Learning in Communications and Networking}, vol.~3,
  pp. 17--29, 2025.

\bibitem{RPAE-DOA-ZXC}
X.~Zhan, Z.~Sun, F.~Shu, Y.~Chen, X.~Cheng, Y.~Wu, Q.~Zhang, Y.~Li, and
  P.~Zhang, ``Rapid phase ambiguity elimination methods for doa estimator via
  hybrid massive mimo receive array,'' \emph{Chinese Journal of Electronics},
  vol.~33, no.~1, pp. 175--184, 2024.

\bibitem{HAD-BJT}
J.~Bai, F.~Shu, F.~Zhou, Q.~Zheng, B.~Xu, B.~Shi, Y.~Chen, W.~Zhang, and
  X.~Wang, ``Co-learning-aided multi-modal-deep-learning framework of passive
  doa estimators for a heterogeneous hybrid massive mimo receiver,'' \emph{IEEE
  Journal of Selected Topics in Signal Processing}, vol.~19, no.~7, pp.
  1448--1460, 2025.

\bibitem{LC-DOA-CYW}
Y.~Chen, X.~Zhan, F.~Shu, Q.~Jie, X.~Cheng, Z.~Zhuang, and J.~Wang, ``Two
  low-complexity doa estimators for massive/ultra-massive mimo receive array,''
  \emph{IEEE Wireless Communications Letters}, vol.~11, no.~11, pp. 2385--2389,
  2022.

\bibitem{DA-UAV-DWK}
J.~Zhang, G.~Lu, L.~Xiang, X.~Ge, and D.~W.~K. Ng, ``Energy minimization for
  uav-aided data collection along a fixed flight path with a directional
  antenna,'' \emph{IEEE Transactions on Communications}, vol.~74, pp. 764--780,
  2026.

\bibitem{2003Antenna}
C.~A. Balanis, ``Antenna theory : analysis and design,'' \emph{IEEE Antennas \&
  Propagation Society Newsletter}, vol.~24, no.~6, pp. 28--29, 2003.

\bibitem{R3}
L.~Zhu, W.~Ma, and R.~Zhang, ``Movable antennas for wireless communication:
  Opportunities and challenges,'' \emph{IEEE Communications Magazine}, vol.~62,
  no.~6, pp. 114--120, 2024.

\bibitem{6D-MA}
X.~Shao, W.~Mei, C.~You, Q.~Wu, B.~Zheng, C.-X. Wang, J.~Li, R.~Zhang,
  R.~Schober, L.~Zhu, W.~Zhuang, and X.~Shen, ``A tutorial on six-dimensional
  movable antenna for 6g networks: Synergizing positionable and rotatable
  antennas,'' \emph{IEEE Communications Surveys \& Tutorials}, vol.~28, pp.
  3666--3709, 2026.

\bibitem{R1}
W.~Ma, L.~Zhu, and R.~Zhang, ``Compressed sensing based channel estimation for
  movable antenna communications,'' \emph{IEEE Communications Letters},
  vol.~27, no.~10, pp. 2747--2751, 2023.

\bibitem{R2}
------, ``Movable antenna enhanced wireless sensing via antenna position
  optimization,'' \emph{IEEE Transactions on Wireless Communications}, vol.~23,
  no.~11, pp. 16\,575--16\,589, 2024.

\bibitem{MA-EB-ZLP}
L.~Zhu, W.~Ma, and R.~Zhang, ``Movable-antenna array enhanced beamforming:
  Achieving full array gain with null steering,'' \emph{IEEE Communications
  Letters}, vol.~27, no.~12, pp. 3340--3344, 2023.

\bibitem{Low-Altitude-MA}
X.~Zhang, W.~Liu, J.~Ren, C.~Wang, H.~Xing, Y.~Shen, and S.~Cui,
  ``Movable-antenna empowered aav-enabled data collection over low-altitude
  wireless networks,'' \emph{IEEE Transactions on Network Science and
  Engineering}, vol.~13, pp. 4506--4523, 2026.

\bibitem{R4}
L.~Zhu, W.~Ma, Z.~Xiao, and R.~Zhang, ``Movable antenna enabled near-field
  communications: Channel modeling and performance optimization,'' \emph{IEEE
  Transactions on Communications}, vol.~73, no.~9, pp. 7240--7256, 2025.

\bibitem{R5}
Y.~Zhang, Y.~Zhang, L.~Zhu, S.~Xiao, W.~Tang, Y.~C. Eldar, and R.~Zhang,
  ``Movable antenna-aided hybrid beamforming for multi-user communications,''
  \emph{IEEE Transactions on Vehicular Technology}, vol.~74, no.~6, pp.
  9899--9903, 2025.

\bibitem{R6}
Y.~Wu, D.~Xu, D.~Wing Kwan~Ng, W.~Gerstacker, and R.~Schober, ``Globally
  optimal movable antenna-enabled multiuser communication: Discrete antenna
  positioning, power consumption, and imperfect csi,'' \emph{IEEE Transactions
  on Communications}, vol.~73, no.~10, pp. 9903--9923, 2025.

\bibitem{R7}
J.~Ding, Z.~Zhou, L.~Zhu, Y.~Zhao, B.~Jiao, and R.~Zhang, ``Energy efficiency
  maximization for movable antenna communication systems,'' \emph{IEEE
  Transactions on Wireless Communications}, vol.~25, pp. 2624--2638, 2026.

\bibitem{FA-NWK}
W.~K. New, K.-K. Wong, H.~Xu, C.~Wang, F.~R. Ghadi, J.~Zhang, J.~Rao, R.~Murch,
  P.~Ramírez-Espinosa, D.~Morales-Jimenez, C.-B. Chae, and K.-F. Tong, ``A
  tutorial on fluid antenna system for 6g networks: Encompassing communication
  theory, optimization methods and hardware designs,'' \emph{IEEE
  Communications Surveys \& Tutorials}, vol.~27, no.~4, pp. 2325--2377, 2025.

\bibitem{FA-KKW2026}
W.~K. New, K.-K. Wong, C.~Wang, C.-B. Chae, R.~Murch, H.~Jafarkhani, and
  Y.~Hao, ``Fluid antenna systems: Redefining reconfigurable wireless
  communications,'' \emph{IEEE Journal on Selected Areas in Communications},
  vol.~44, pp. 1013--1044, 2026.

\bibitem{RIS-HCW}
C.~Huang, R.~Mo, and C.~Yuen, ``Reconfigurable intelligent surface assisted
  multiuser miso systems exploiting deep reinforcement learning,'' \emph{IEEE
  Journal on Selected Areas in Communications}, vol.~38, no.~8, pp. 1839--1850,
  2020.

\bibitem{DM-IRS-SF}
F.~Shu, Y.~Teng, J.~Li, M.~Huang, W.~Shi, J.~Li, Y.~Wu, and J.~Wang, ``Enhanced
  secrecy rate maximization for directional modulation networks via irs,''
  \emph{IEEE Transactions on Communications}, vol.~69, no.~12, pp. 8388--8401,
  2021.

\bibitem{IRS-Sensing-WQQ}
Z.~Zhang, Q.~Wu, W.~Chen, Y.~Zhu, Z.~Zheng, Y.~Gao, and Q.~Wu, ``Irs-aided
  secure sensing for surveillance area coverage: Framework and algorithm
  design,'' \emph{IEEE Journal on Selected Areas in Communications}, pp. 1--1,
  2026.

\bibitem{IRS-DM-DRE}
D.~Rongen, S.~Feng, L.~Yongzhao, T.~Yanqun, L.~Jun, W.~Yongpeng, and
  W.~Jiangzhou, ``Joint power allocation and beamforming for active irs-aided
  secure directional modulation network,'' \emph{Science China Information
  Sciences}, vol.~68, p. 222301, 2025.

\bibitem{DoF-JJB}
\BIBentryALTinterwordspacing
J.~Jiang, F.~Shu, X.~Wang, K.~Yang, C.~Shen, Q.~Zhang, D.~Wang, and J.~Wang,
  ``Dof analysis and beamforming design for active irs-aided multi-user mimo
  wireless communication in low-rank channels,'' \emph{Tsinghua Science and
  Technology}, 2025. [Online]. Available:
  \url{https://www.sciopen.com/article/10.26599/TST.2025.9010084}
\BIBentrySTDinterwordspacing

\bibitem{RA-2ZBX}
X.~Xiong, B.~Zheng, W.~Wu, W.~Zhu, M.~Wen, S.~Lin, and Y.~Zeng, ``Intelligent
  rotatable antenna for integrated sensing, communication, and computation:
  Challenges and opportunities,'' \emph{IEEE Wireless Communications}, vol.~33,
  no.~1, pp. 173--180, 2026.

\bibitem{RA-ZBX}
B.~Zheng, T.~Ma, C.~You, J.~Tang, R.~Schober, and R.~Zhang, ``Rotatable antenna
  enabled wireless communication and sensing: Opportunities and challenges,''
  \emph{IEEE Wireless Communications}, pp. 1--8, 2025.

\bibitem{RAA-UAV}
X.~Zhang, L.~Xiang, J.~Wang, X.~Gao, D.~W.~K. Ng, and R.~Schober, ``Rotatable
  antenna array enabled uav mmwave massive mimo communication,'' \emph{IEEE
  Transactions on Communications}, vol.~74, pp. 1219--1236, 2026.

\bibitem{Jiang2025RotatableAD}
J.~Jiang, F.~Shu, B.~Deng, M.~Li, J.~Bai, Y.~Wang, C.~Pan, and J.~Wang,
  ``Rotatable antenna-array-enhanced direction-sensing for low-altitude
  communication network: Method and performance,'' 2025.

\bibitem{Gurney1961}
R.~Gurney, \emph{The Printing of Mathematics}.\hskip 1em plus 0.5em minus
  0.4em\relax Toronto: University of Toronto Press, 1961, pp. 127--134.

\bibitem{2009Classical}
Tuncer and T.~Engin, ``Classical and modern direction-of-arrival estimation,''
  \emph{Academic Press}, pp. 125--160, 2009.

\bibitem{LC-CRLB}
A.~Wang, L.~Liu, and J.~Zhang, ``Low complexity direction of arrival (doa)
  estimation for 2d massive mimo systems,'' in \emph{2012 IEEE Globecom
  Workshops}, 2012, pp. 703--707.

\bibitem{2006Fundamentals}
\emph{Fundamentals of Statistical Signal Processing}.\hskip 1em plus 0.5em
  minus 0.4em\relax London: Springer London, 2006, pp. 83--182.

\bibitem{1994Fundamentals}
S.~K. Sengijpta, ``Fundamentals of statistical signal processing: Estimation
  theory,'' \emph{Control Engineering Practice}, vol.~37, no.~4, pp. 465--466,
  1994.

\end{thebibliography}

\end{document}